\documentclass[runningheads]{llncs}
\usepackage{graphicx}
\usepackage{tabularx}
\usepackage{color,soul}
\usepackage{rotating}

\newcommand*{\myfont}{\fontfamily{pcr}\selectfont}

\usepackage{floatrow}
\newfloatcommand{capbtabbox}{table}[][\FBwidth]

\begin{document}
%

\title{Can We Assess Mental Health through Social Media and Smart Devices? Addressing Bias in Methodology and Evaluation
}
\titlerunning{Can We Assess Mental Health through Social Media and Smart Devices?}
%

\author{Adam Tsakalidis\inst{1,2} \and
Maria Liakata\inst{1,2} \and
Theo Damoulas\inst{1,2}\and
Alexandra I. Cristea\inst{1,3}}
\authorrunning{A. Tsakalidis et al.}
%
\institute{University of Warwick, UK\\ \email{\{a.tsakalidis, m.liakata, t.damoulas, a.i.cristea\}@warwick.ac.uk}
\and
The Alan Turing Institute, UK\\
\and
Durham University, UK}
%

\maketitle              

\begin{abstract}
Predicting mental health from smartphone and social media data on a longitudinal basis has recently attracted great interest, with very promising results being reported across many studies \cite{canzian2015trajectories,jaques2015predicting,likamwa2013moodscope,tsakalidis2016combining}. Such approaches have the potential to revolutionise mental health assessment, if their development and evaluation follows a real world deployment setting. In this work we take a closer look at state-of-the-art approaches, using different mental health datasets and indicators, different feature sources and multiple simulations, in order to assess their ability to generalise. We demonstrate that under a pragmatic evaluation framework, none of the approaches deliver or even approach the reported performances. In fact, we show that current state-of-the-art approaches can barely outperform the most na\"ive baselines in the real-world setting, posing serious questions not only about their deployment ability, but also about the contribution of the derived features for the mental health assessment task and how to make better use of such data in the future.

\keywords{mental health \and bias \and evaluation \and wellbeing \and natural language processing \and smartphones \and sensors \and social media}
\end{abstract}

\section{Introduction}

Establishing the right indicators of mental well-being is a grand challenge posed by the World Health Organisation \cite{herrman2005promoting}.  Poor mental health is highly correlated with low motivation, lack of satisfaction, low productivity and a negative economic impact \cite{olesen2012economic}.  The current approach is to combine census data at the population level \cite{oecd2013howslife}, thus failing to capture well-being on an individual basis. The latter is only possible via self-reporting on the basis of established psychological scales, which are hard to acquire consistently on a longitudinal basis, and they capture long-term aggregates instead of the current state of the individual.


The widespread use of smart-phones and social media offers new ways of assessing mental well-being, and recent research \cite{bogomolov2014pervasive,bogomolov2013happiness,canzian2015trajectories,farhan2016behavior,jaques2015predicting,jaques2015multi,likamwa2013moodscope,ma2012daily,servia2017mobile,suhara2017deepmood,tsakalidis2016combining} has started exploring the effectiveness of these modalities for automatically assessing the mental health of a subject, reporting very high accuracy. What is typically done in these studies is to use features based on the subjects' smart phone logs and social media, to predict some self-reported mental health index (e.g., ``wellbeing'', ``depression'' and others), which is provided either on a Likert scale or on the basis of a psychological questionnaire (e.g., PHQ-8 \cite{kroenke2009phq},  PANAS \cite{watson1988development}, WEMWBS \cite{tennant2007warwick} and others).

Most of these studies are longitudinal, where data about individuals is collected over a period of time and predictions of mental health are made over a sliding time window.  Having such longitudinal studies is highly desirable, as it can allow fine-grained monitoring of mental health. However, a crucial question is \textit{what constitutes an appropriate evaluation framework}, in order for such approaches to be employable in a real world setting. Generalisation to previously unobserved users can only be assessed via leave-N-users-out cross-validation setups, where typically, N is equal to one (\textbf{LOUOCV}, see Table~\ref{tab:intro1}). However, due to the small number of subjects that are available, such generalisation is hard to achieve by any approach \cite{likamwa2013moodscope}. Alternatively, personalised models \cite{canzian2015trajectories,likamwa2013moodscope} for every individual can be evaluated via a within-subject, leave-N-instances-out cross-validation (for N=1,  \textbf{LOIOCV}), where an instance for a user $u$ at time $i$ is defined as a \{$X_{ui}$, $y_{ui}$\} tuple of \{features(u, i), mental-health-score(u, i)\}. In a real world setting, a \textit{LOIOCV} model is trained on some user-specific instances, aiming to predict her mental health state at some future time points. Again however, the limited number of instances for every user make such models unable to generalize well. In order to overcome these issues, previous work \cite{bogomolov2013happiness,farhan2016behavior,jaques2015predicting,jaques2015multi,servia2017mobile,tsakalidis2016combining} has combined the instances \{$X_{u_ji}$, $y_{u_ji}$\} from different individuals $u_j$ and performed evaluation using randomised cross validation (\textbf{MIXED}). While such approaches can attain optimistic performance, the corresponding models fail to generalise to the general population and also fail to ensure effective personalised assessment of the mental health state of a single individual.

\begin{table}[ht]
\centering
\caption{Summary of the three evaluation frameworks.}
\label{tab:intro1}
\resizebox{\columnwidth}{!}{%
\begin{tabular}{|p{1.05cm}|p{3.5cm}|p{5cm}|p{5.6cm}|} \hline
       & \hspace{.9cm}\textbf{LOUOCV} & 
       \hspace{1.6cm}\textbf{LOIOCV} & 
       \hspace{2cm}\textbf{MIXED}\\ \hline
       

\textbf{Real world aim} & Build a model $m$ that generalises to a previously unseen user $u$ & Build a personalised model $m_u$ per user $u$ that generalises on $u$, given some manual input by $u$ & Build a model $m$ that generalises to new instances of a specific pool of previously seen users \\\hline


\textbf{Train} & \{\{$X_{\not u i}$, $y_{\not u i}$\}\} & \{\{$X_{u \not i}$, $y_{u \not i}$\}\} & \{\{$X_{u_0 \not i}$, $y_{u_0 \not i}$\}, ..., \{$X_{u_n \not i}$, $y_{u_n \not i}$\}\}
\\\hline 

\textbf{Test}& \{$X_{ui}$, $y_{ui}$\} & \{$X_{u i}$, $y_{u i}$\}&\{\{$X_{u_0 i}$, $y_{u_0 i}$\}, ..., \{$X_{u_n i}$, $y_{u_n  i}$\}\}\\\hline


\textbf{\newline Limits} & Few users for training and evaluation & Few instances per user for training and evaluation & Cannot ensure generalisation neither over new users nor in a personalised way \\ \hline

\end{tabular}
}
\end{table}

In this paper we demonstrate the challenges that current state-of-the-art models face, when tested in a real-world setting. We work on two longitudinal datasets with four mental health targets, using different features derived from a wide range of heterogeneous sources. Following the state-of-the-art experimental methods and evaluation settings, we achieve very promising results, regardless of the features we employ and the mental health target we aim to predict. However, when tested under a pragmatic setting, the performance of these models drops heavily, \textit{failing to outperform the most na\"ive -- from a modelling perspective -- baselines}: majority voting, random classifiers, models trained on the identity of the user, etc. This poses serious questions about the contribution of the features derived from social media, smartphones and sensors for the task of automatically assessing well-being on a longitudinal basis. Our goal is to flesh out, study and discuss such limitations through extensive experimentation across multiple settings, and to propose a pragmatic evaluation and model-building framework for future research in this domain.


\section{Related Work}

Research in \textit{assessing mental health on a longitudinal basis} aims to make use of relevant features extracted from various modalities, in order to train models for automatically predicting a user's mental state (target), either in a classification or a regression manner \cite{bogomolov2014pervasive,bogomolov2013happiness,canzian2015trajectories,jaques2015predicting,jaques2015multi,likamwa2013moodscope,tsakalidis2016combining}. Examples of state-of-the-art work in this domain are listed in Table \ref{tab:litrev2}, along with the number of subjects that was used and the method upon which evaluation took place. Most approaches have used the \textit{``MIXED"} approach to evaluate models \cite{bogomolov2014pervasive,bogomolov2013happiness,farhan2016behavior,jaques2015predicting,jaques2015multi,servia2017mobile,tsakalidis2016combining}, which, as we will show, is vulnerable to bias, due to the danger of recognising the user in the test set and thus simply inferring her average mood score. \textit{LOIOCV} approaches that have not ensured that their train/test sets are independent are also vulnerable to bias in a realistic setting \cite{canzian2015trajectories,likamwa2013moodscope}. From the works listed in Table \ref{tab:litrev2}, only Suhara et al. \cite{suhara2017deepmood} achieves unbiased results with respect to model generalisability; however, the features employed for their prediction task are derived from self-reported questionnaires of the subjects and not by automatic means.

\begin{table}[t!]
\caption{Works on predicting mental health in a longitudinal manner.}
\label{tab:litrev2}

\resizebox{\textwidth}{!}{
\begin{tabular}{|p{3.48cm}|p{3.2cm}|p{6cm}|c|r|p{1.4cm}|}
\hline
\hspace{1.2cm}\textbf{Work}&\hspace{1.1cm}\textbf{Target}&\hspace{2.2cm}\textbf{Modalities}&\textbf{Type}&\textbf{Size}&\hspace{.42cm}\textbf{Eval}\\ \hline \hline

Ma et al. \cite{ma2012daily}&Tiredness, Tensity, Displeasure (1-5)&location, accelerometer,  sms, calls&Class.&15&N/A\\ \hline

Bogomolov et al. \cite{bogomolov2013happiness}&Happiness (1-7)&weather, calls, bluetooth, sms, Big Five&Class.&117&MIXED\\ \hline

LiKamWa et al. \cite{likamwa2013moodscope}& Activeness, Pleasure (1-5)&email/phone/sms contacts, location, apps, websites&Regr.&32&LOIOCV LOUOCV\\ \hline

Bogomolov et al. \cite{bogomolov2014pervasive}&Stress (1-7)&weather, calls, bluetooth, sms, Big Five&Class.&117&MIXED\\ \hline

Canzian and Musolesi \cite{canzian2015trajectories}&PHQ-8&GPS&Class.&48&LOIOCV\\ \hline



Jaques et al. \cite{jaques2015predicting,jaques2015multi}&Happiness \cite{jaques2015predicting},  \{Happiness, Health, Stress, Energy, Alertness\} (0-100) \cite{jaques2015multi} &electrodermal activity, calls,  accelerometer, sms, surveys, phone usage, locations &Class.&68&MIXED\\ \hline

Tsakalidis et al. \cite{tsakalidis2016combining}&PANAS, WEMWBS &social media, calls, sms, locations, headphones, charger,  screen/ringer mode, wifi &Regr.&19&MIXED\\ \hline 

Farhan et al. \cite{farhan2016behavior}&PHQ-9&GPS, PHQ-9 scores&Class.&79&MIXED\\\hline

Wang et al. \cite{wang2016crosscheck}&Positive, Negative (0-15),  Positive-Negative&GPS, calls, accelerometer, microphone, light sensor, sms, apps, phone locked&Regr.&21&LOIOCV LOUOCV\\ \hline

Servia-Rodriguez et al. \cite{servia2017mobile}&Positive/Negative, Alert/Sleepy&microphone, accelerometer, calls, sms&Class.&726&MIXED\\ \hline

Suhara et al. \cite{suhara2017deepmood}& Mood (binary)&daily surveys&Class.&2,382&\begin{tabular}[c]{@{}l@{}}LNUOCV\\ \end{tabular}\\ \hline

\end{tabular}
}
\end{table}


\section{Problem Statement}
\label{sec:3}
We first describe three  major problems stemming from unrealistic construction and evaluation of mental health assessment models and then we briefly present the state-of-the-art in each case, which we followed in our experiments.  

\begin{itemize}
    \item[\textbf{\myfont P1}] \textbf{Training on past values of the target variable}: This issue arises when the past $N$ mood scores of a user are required to predict his/her next mood score in an autoregressive manner. Since such an approach would require the previous N scores of past mood forms, it would limit its ability to generalise without the need of manual user input in a continuous basis. This makes it impractical for a real-world scenario. Most importantly, it is difficult to measure the contribution of the features towards the prediction task, unless the model is evaluated using target feature ablation. For demonstration purposes, we have followed the experimental setup by LiKamWa et al. \cite{likamwa2013moodscope}, which is one of the leading works in this field.
    
    \item[\textbf{\myfont P2}] \textbf{Inferring test set labels}: When training personalised models (\textit{LOIOCV}) in a longitudinal study, it is important to make sure that there are no overlapping instances across consecutive time windows. Some past works have extracted features \{$f(t-N)$, ..., $f(t)$\} over $N$ days, in order to predict the $score_t$ on day $N+1$ \cite{canzian2015trajectories,likamwa2013moodscope}. Such approaches are biased if there are overlapping days of train/test data. To illustrate this problem we have followed the approach by Canzian and Musolesi \cite{canzian2015trajectories}, as one of the pioneering works on predicting depression with GPS traces, on a longitudinal basis.
    
    \item[\textbf{\myfont P3}] \textbf{Predicting users instead of mood scores}:  Most approaches merge all the instances from different subjects, in an attempt to build user-agnostic models in a randomised cross-validation framework \cite{bogomolov2013happiness,jaques2015predicting,jaques2015multi,tsakalidis2016combining}. This is problematic, especially when dealing with a small number of subjects, whose behaviour (as captured through their data) and mental health scores differ on an individual basis. Such approaches are in danger of ``predicting'' the user in the test set, since her (test set) features might be highly correlated with her features in the training set, and thus infer her average well-being score, based on the corresponding observations of the training set. Such approaches cannot guarantee that they will generalise on either a population-wide (\textit{LOUOCV}) or a personalised (\textit{LOIOCV}) level. In order to examine this effect in both a regression and a classification setting, we have followed the experimental framework by Tsakalidis et al. \cite{tsakalidis2016combining} and Jaques et al. \cite{jaques2015predicting}. 
\end{itemize}



\subsection{{\myfont P1}: Training on past values of the target (LOIOCV, LOUOCV)} 
\label{sec:likamwa}
LiKamWa et al. \cite{likamwa2013moodscope} collected smartphone data from $32$ subjects over a period of two months. The subjects were asked to self-report their ``pleasure'' and ``activeness'' scores at least four times a day, following a Likert scale (1 to 5), and the average daily scores served as the two targets. The authors aggregated various features on social interactions (e.g., number of emails sent to frequently interacting contacts) and routine activities (e.g., browsing and location history)  derived from the smartphones of the participants. These features were extracted over a period of three days, along with the two most recent scores on activeness and pleasure. The issue that naturally arises is that such a method cannot generalise to new subjects in the \textit{LOUOCV} setup, as it requires their last two days of self-assessed scores. Moreover, in the \textit{LOIOCV} setup, the approach is 
limited in a real world setting, since it requires the previous mental health scores by the subject to provide an estimate of her current state. Even in this case though, the feature extraction should be based on past information only -- under \textit{LOIOCV} in \cite{likamwa2013moodscope}, the current mood score we aim at predicting is also used as a feature in the (time-wise) subsequent two instances of the training data.

Experiments in \cite{likamwa2013moodscope} are conducted under \textit{LOIOCV} and \textit{LOUOCV}, using Multiple Linear Regression (LR) with Sequential Feature Selection (in \textit{LOUOCV}, the past two pairs of target labels of the test user are still used as features). In order to better examine the effectiveness of the features for the task, the same model can be tested without any ground-truth data as input. Nevertheless, a simplistic model predicting the per-subject average outperforms their LR in the \textit{LOUOCV} approach, which poses the question of whether the smartphone-derived features can be used effectively to create a generalisable model that can assess the mental health of unobserved users. Finally, the same model tested in the \textit{LOIOCV} setup achieves the lowest error; however, this is trained not only on target scores overlapping with the test set, but also on features derived over a period of three days, introducing further potential bias, as discussed in the following.

\subsection{{\myfont P2}: Inferring Test Labels (LOIOCV)}
\label{sec:canzian}
Canzian and Musolesi \cite{canzian2015trajectories} extracted mobility metrics from $28$ subjects to predict their depressive state, as derived from their daily self-reported PHQ-8 questionnaires. A 14-day moving average filter is first applied to the PHQ-8 scores and the mean value of the same day (e.g. Monday) is subtracted from the normalised scores, to avoid cyclic trends. This normalisation results into making the target score $s_t$ on day $t$ dependent on the past \{$s_{t-14}, ..., s_{t-1}$\} scores. The normalised PHQ-8 scores are then converted into two classes, with the instances deviating more than one standard deviation above the mean score of a subject being assigned to the class ``1'' (``0'', otherwise). The features are extracted over various time windows (looking at $T_{HIST}=\{0,...,14\}$ days before the completion of a mood form) and personalised model learning and evaluation are performed for every $T_{HIST}$ separately, using a \textit{LOIOCV} framework.

What is notable is that the results improve significantly when features are extracted from a wider $T_{HIST}$ window. This could imply that the depressive state of an individual can be detected with a high accuracy if we look back at her history. However, by training and testing a model on instances whose features are derived from the same days, there is a high risk of over-fitting the model to the timestamp of the day in which the mood form was completed. In the worst-case scenario, there will be an instance in the train set whose features (e.g. total covered distance) are derived from the 14 days, 13 of which will also be used for the instance in the test set. Additionally, the target values of these two instances will also be highly correlated due to the moving average filter, making the task artificially easy for large $T_{HIST}$ and not applicable in a real-world setting.

While we focus on the approach in \cite{canzian2015trajectories}, a similar approach with respect to feature extraction was also followed in LiKamWa et al. \cite{likamwa2013moodscope} and Bogomolov et al. \cite{bogomolov2013happiness}, extracting features from the past 2 and 2 to 
5 days, respectively.

\subsection{{\myfont P3}: Predicting Users (LOUOCV)}
\label{sec:tsakalidis}
Tsakalidis et al. \cite{tsakalidis2016combining} monitored the behaviour of $19$ individuals over four months. The subjects were asked to complete two psychological scales \cite{tennant2007warwick,watson1988development} on a daily basis, leading to three target scores (positive, negative, mental well-being); various features from smartphones (e.g., time spent on the preferred locations) and textual features (e.g., ngrams) were extracted passively over the 24 hours preceding a mood form timestamp. Model training and evaluation was performed in a randomised (\textit{MIXED}) cross-validation setup, leading to high accuracy ($R^2=0.76$). However, a case demonstrating the potential user bias is when the models are trained on the textual sources: initially the highest $R^2$ ($0.22$) is achieved when a  model is applied to the mental-wellbeing target; by normalising the textual features on a per-user basis, the $R^2$ increases to $0.65$. While this is likely to happen because the  vocabulary used by different users is normalised, there is also the danger of over-fitting the trained model to the identity of the user. To examine this potential, the \textit{LOIOCV/LOUOCV} setups need to be studied alongside the \textit{MIXED} validation approach, with and without the per-user feature normalisation step.

A similar issue is encountered in Jaques et al. \cite{jaques2015predicting} who monitored $68$ subjects over a period of a month. Four types of features were extracted from survey and smart devices carried by subjects. Self-reported scores on a daily basis served as the ground truth. The authors labelled the instances with the top $30\%$ of all the scores as ``happy'' and the lowest $30\%$ as ``sad'' and randomly separated them into training, validation and test sets, leading to the same user bias issue. Since different users exhibit different mood scores on average \cite{tsakalidis2016combining}, by selecting instances from the top and bottom scores, one might end up separating users and convert the mood prediction task into a user identification one. A more suitable task could have been to try to predict the highest and lowest scores of every individual separately, either in a \textit{LOIOCV} or in a \textit{LOUOCV} setup.

While we focus on the works of  Tsakalidis et al. \cite{tsakalidis2016combining} and Jaques et al. \cite{jaques2015predicting}, similar experimental setups were also followed in \cite{jaques2015multi}, using the median of scores to separate the instances and performing five-fold cross-validation, and by Bogomolov et al. in \cite{bogomolov2013happiness}, working on a user-agnostic validation setting on 117 subjects to predict their happiness levels, and in \cite{bogomolov2014pervasive}, for the stress level classification task.
\section{Experiments}
\subsection{Datasets}
By definition, the aforementioned issues are feature-, dataset- and target-independent (albeit the magnitude of the effects may vary). To illustrate this, we run a series of experiments employing two datasets, with different feature sources and four different mental health targets.

\vspace{.125cm}
\noindent\textbf{Dataset 1}: We employed the dataset obtained by Tsakalidis et al. \cite{tsakalidis2016combining}, a pioneering dataset which contains a mix of longitudinal textual and mobile phone usage data for 30 subjects. From a textual perspective, this dataset consists of social media posts (1,854/5,167 facebook/twitter posts) and private messages (64,221/132/47,043 facebook/twitter/ SMS messages) sent by the subjects. For our ground truth, we use the \{positive, negative, mental well-being\} mood scores (in the ranges of [10-50], [10-50], [14-70], respectively) derived from self-assessed psychological scales during the study period.

\vspace{.125cm}
\noindent\textbf{Dataset 2}: We employed the StudentLife dataset \cite{wang2014studentlife}, which contains a wealth of information derived from the smartphones of 48 students during a 10-week period. Such information includes samples of the detected activity of the subject, timestamps of detected conversations, audio mode of the smartphone, status of the smartphone (e.g., charging, locked), etc. For our target, we used the self-reported stress levels of the students (range [0-4]), which were provided several times a day. For the approach in LiKamWa et al. \cite{likamwa2013moodscope}, we considered the average daily stress level of a student as our ground-truth, as in the original paper; for the rest, we used all of the stress scores and extracted features based on some time interval preceding their completion, as described next, in 4.3\footnote{For {\myfont P3}, this creates the {\myfont P2} cross-correlation issue in the \textit{MIXED}/\textit{LOIOCV} settings. For this reason, we  ran the experiments by considering only the last entered score in a given day  as our target. We did not witness any major differences that would alter our conclusions.}.

\subsection{Task Description}
\label{sec:taskdesc}
We studied the major issues in the following experimental settings (see Table \ref{table:experiments}): 

\vspace{.1cm}
    \noindent \textbf{{\myfont P1}: Using Past Labels}:
    We followed the experimental setting in \cite{likamwa2013moodscope} (see section \ref{sec:likamwa}): we treated our task as a regression problem and used Mean Squared Error (MSE) and classification accuracy\footnote{Accuracy is defined in \cite{likamwa2013moodscope} as follows: 5 classes are assumed (e.g., [0, ..., 4]) and the squared error $e$ between the centre of a class halfway towards the next class is calculated (e.g., 0.25). If the squared error of a test instance is smaller than $e$, then it is considered as having been classified correctly.} for evaluation. We trained a Linear Regression (LR) model and performed feature selection using Sequential Feature Selection under the \textit{LOIOCV} and \textit{LOUOCV} setups; feature extraction is performed over the previous 3 days preceding the completion of a mood form. For comparison, we use the same baselines as in \cite{likamwa2013moodscope}: Model A always predicts  the average mood score for a certain user (\texttt{AVG}); Model B predicts the last entered scores (\texttt{LAST}); Model C makes a prediction  using  the  LR model trained on the ground-truth features only (\texttt{-feat}). We also include Model D, trained on non-target features only (\texttt{-mood}) in an unbiased \textit{LOUOCV} setting.

    \vspace{.1cm}
    \noindent \textbf{{\myfont P2}: Inferring Test Labels}: We followed the experimental setting presented in \cite{canzian2015trajectories}. We process our ground-truth in the same way as the original paper (see section \ref{sec:canzian}) and thus treat our task as a binary classification problem. We use an SVM$_{RBF}$ classifier, using grid search for parameter optimisation, and perform evaluation using specificity and sensitivity. We run experiments in the \textit{LOIOCV} and \textit{LOUOCV} settings, performing feature extraction at different time windows ($T_{HIST}=\{1, ..., 14\}$). In order to better demonstrate the problem that arises here, we use the previous label classifier (\texttt{LAST}) and the SVM classifier to which we feed only the mood timestamp as a feature (\texttt{DATE}) for comparison. Finally, we replace our features with completely random data and train the same SVM with $T_{HIST}=14$ by keeping the same ground truth, performing 100 experiments and reporting averages of sensitivity and specificity  (\texttt{RAND}). 
    
    \vspace{.1cm}
    \noindent \textbf{{\myfont P3}: Predicting Users}: We followed the evaluation settings of two past works (see section \ref{sec:tsakalidis}), with the only difference being the use of 5-fold CV instead of a train/dev/test split that was used in \cite{jaques2015predicting}. The features of every instance are extracted from the past day before the completion of a mood form. In \textbf{Experiment 1} we follow the setup in \cite{tsakalidis2016combining}: we perform 5-fold CV (\textit{MIXED}) using SVM (SVR$_{RBF}$) and evaluate performance based on $R^2$ and $RMSE$. We compare the performance when tested under the \textit{LOIOCV}/\textit{LOUOCV} setups, with and without the per-user feature normalisation step. We also compare the performance of the \textit{MIXED} setting, when our model is trained on the one-hot-encoded user id only. In \textbf{Experiment 2} we follow the setup in \cite{jaques2015predicting}: we label the instances as ``high'' (``low''), if they belong to the top-30\% (bottom-30\%) of mood score values (``UNIQ'' -- for ``unique'' -- setup). We train an SVM classifier in 5-fold CV using accuracy for evaluation and compare performance in the \textit{LOIOCV} and \textit{LOUOCV} settings. In order to further examine user bias, we perform the same experiments, this time by labelling the instances on a per-user basis (``PERS'' -- for ``personalised'' -- setup), aiming to predict the per-user high/low mood days\footnote{In cases where the lowest of the top-30\% scores ($s$) was equal to the highest of the bottom-30\% scores, we excluded the instances with score $s$.
    }.

\begin{table}[h]
\caption{Summary of experiments. The highlighted settings indicate the settings used in the original papers; ``Period'' indicates the period before each mood form completion during which the features were extracted.}
\label{table:experiments}
\centering
\resizebox{.9\columnwidth}{!}{%
\begin{tabular}{|l||c|c|c|}
\hline
Issue & {\myfont P1}: Training on past labels & {\myfont P2}: Inferring test labels &{\myfont P3}: Predicting users\\ \hline \hline

Setting & \textbf{LOIOCV},\textbf{LOUOCV} & \textbf{LOIOCV}, LOUOCV & \textbf{MIXED}, LOIOCV, LOUOCV\\\hline

Task&Regr.&Class.&Regr. (E1); Class. (E2)\\\hline

Metrics &MSE, accuracy& sensitivity, specificity& $R^2$, RMSE (E1); accuracy (E2)\\\hline
Period&Past 3 days &Past \{1,...,14\} days& Past day\\\hline

Model & LR$_{sfs}$ & SVM$_{rbf}$ & SVR$_{rbf}$; SVM$_{rbf}$\\\hline

Baselines&\texttt{AVG}, \texttt{LAST}, \texttt{-feat}, \texttt{-mood}&\texttt{LAST}, \texttt{DATE},\texttt{RAND}&model trained on user id\\
\hline
\end{tabular}}
\end{table}

\subsection{Features}
\label{4.3}
For \textit{Dataset 1}, we first defined a ``user snippet'' as the concatenation of all texts generated by a user within a set time interval, such that the maximum time difference between two consecutive document timestamps is less than 20 minutes. We performed some standard noise reduction steps (converted text to lowercase, replaced URLs/user mentions and performed language identification\footnote{\url{https://pypi.python.org/pypi/langid}} and tokenisation \cite{gimpel2011part}). Given a mood form and a set of snippets produced by a user before the completion of a mood form, we extracted some commonly used feature sets for every snippet written in English \cite{tsakalidis2016combining}, which were used in all experiments. To ensure sufficient data density, we excluded users for whom we had overall fewer than $25$ snippets on the days before the completion of the mood form or fewer than $40$ mood forms overall, leading to $27$ users and $2,368$ mood forms.  For \textit{Dataset 2}, we extracted the features presented in Table \ref{table:features}. We only kept the users that had at least 10 self-reported stress questionnaires, leading to $44$ users and $2,146$ instances. For our random experiments used in {\myfont P2}, in Dataset 1 we replaced the text representation of every snippet with random noise ($\mu=0, \sigma=1$) of the same feature dimensionality; in Dataset 2, we replaced the actual inferred value of every activity/audio sample with a random inference class; we also replaced each of the detected conversation samples and samples detected in a dark environment/locked/charging, with a random number ($<$100, uniformly distributed) indicating the number of pseudo-detected samples.

\begin{table}[h]
\caption{Features that were used in our experiments in Datasets 1, 2 (left, right).}
\label{table:features}
\centering
\resizebox{\columnwidth}{!}{%
\begin{tabular}{|p{10.35cm}||p{5.65cm}|}
\hline

\textit{(a)} \textbf{duration} of the snippet; \textit{(b)} binary \textbf{ngrams} ($n=1, 2$); \textit{(c)} cosine similarity between the words of the document and the 200 \textbf{topics} obtained by \cite{income}; \textit{(d)} functions over  \textbf{word} \textbf{embeddings} dimensions \cite{tang} (mean, max, min, median, stdev, 1st/3rd quartile); \textit{(e)} \textbf{lexicons} \cite{opinionlexicon,nrcuni1,nrchashtagemotion,msol,afinn,saif14}: for lexicons providing binary values (pos/neg), we counted the number of ngrams matching each class and for those with score values, we used the counts and the total summation of the corresponding scores. \textit{(f)} \textbf{number} of Facebook posts/messages/images, Twitter posts/messages, SMS, number of tokens/messages/posts in the snippet.
&\textit{(a)} percentage of collected samples for each \textbf{activity} (stationary, walking, running, unknown) and \textit{(b)} \textbf{audio} mode (silence, voice, noise, unknown); \textit{(c)} number and total duration of detected \textbf{conversations}; number of samples and total duration of the time during which the phone was \textit{(d)} in a \textbf{dark} \textbf{environment}, \textit{(e)} \textbf{locked} and \textit{(f)} \textbf{charging}.\\\hline
\end{tabular}}
\end{table}

\section{Results}

\subsection{{\myfont P1}: Using Past Labels}
Table \ref{tab:moodscope_cross} presents the results on the basis of the methodology by LiKamWa et al. \cite{likamwa2013moodscope}, along with the average scores reported in \cite{likamwa2013moodscope} -- note that the range of the mood scores varies on a per-target basis; hence, the reported results of different models should be compared among each other when tested on the \textit{same} target.

\begin{table}[t!]
\resizebox{.8\textwidth}{!}{
\begin{tabular}{ |l||r|c||c|c||c|c||c|c||c|c|}\hline
 & \multicolumn{2}{c||}{positive} &
 \multicolumn{2}{c||}{negative} & 
 \multicolumn{2}{c||}{wellbeing}&
 \multicolumn{2}{c||}{stress}&
 \multicolumn{2}{c|}{ \cite{likamwa2013moodscope}}\\
 \hline
 &MSE&acc&MSE&acc&MSE&acc&MSE&acc&MSE & acc\\\hline
LOIOCV&15.96&84.5&11.64&87.1&20.94&89.0&  1.07 & 47.3 &0.08&93.0\\
LOUOCV& 36.77 & 63.4 & 31.99 & 68.3 & 51.08 & 72.8& 0.81& 45.4  &0.29&66.5\\
 A (\texttt{AVG})&29.89 & 71.8&  27.80& 73.1 & 41.14 & 78.9& 0.70 & 51.6 &0.24&73.5\\ 
 B (\texttt{LAST})&43.44&60.4&38.22&63.2&55.73&71.6& 1.15 & 51.5  &0.34&63.0\\
 C (\texttt{-feat})&33.40 & 67.2 & 28.60 & 72.3 & 45.66 & 76.6& 0.81& 49.8   &0.27&70.5\\
 D (\texttt{-mood}) &113.30& 30.9 & 75.27 & 44.5 & 138.67 & 42.5& 1.08& 44.4   &N/A&N/A\\
\hline
\end{tabular}
}
\caption{{\myfont P1}: Results following the approach in \cite{likamwa2013moodscope}. 
}
\label{tab:moodscope_cross}
\end{table}

As in \cite{likamwa2013moodscope}, always predicting the average score (\texttt{AVG}) for an unseen user performs better than applying a LR model trained on other users in a \textit{LOUOCV} setting. If the same LR model used in \textit{LOUOCV} is trained without using the previously self-reported ground-truth scores (Model D, \texttt{-mood}), its performance drops further. This showcases that personalised models are needed for more accurate mental health assessment (note that the \texttt{AVG} baseline is, in fact, a personalised baseline) 
and that there is no evidence that we can employ effective models in real-world applications to predict the mental health of previously unseen individuals, based on this setting.

The accuracy of LR under \textit{LOIOCV} is higher, except for the ``stress'' target, where the performance is comparable to \textit{LOUOCV} and lower than the \texttt{AVG} baseline. However, the problem in \textit{LOIOCV} is the fact that the features are extracted based on the past three days, thus creating a temporal cross-correlation in our input space. If a similar  correlation exists in the output space (target), then we end up in danger of overfitting our model to the training examples that are temporally close to the test instance. This type of bias is essentially present if we force a temporal correlation in the output space, as studied next.

\subsection{{\myfont P2}: Inferring Test Labels}

\begin{figure}[ht]
  \includegraphics[width=.23\columnwidth, trim={1.22cm 0.82cm 2.0cm 1.1cm},clip]{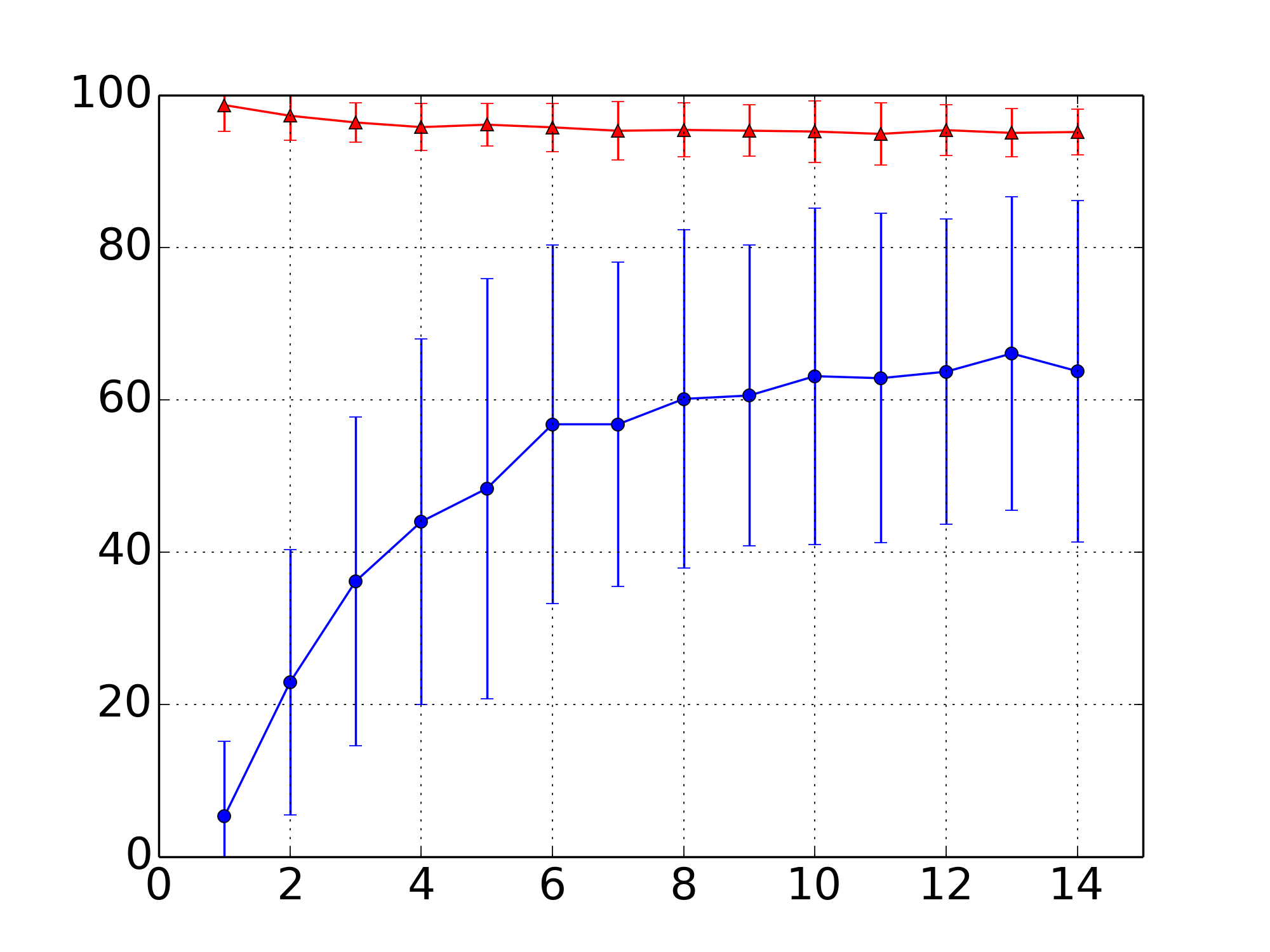}
  \includegraphics[width=.23\columnwidth, trim={1.22cm 0.82cm 2.0cm 1.1cm},clip]{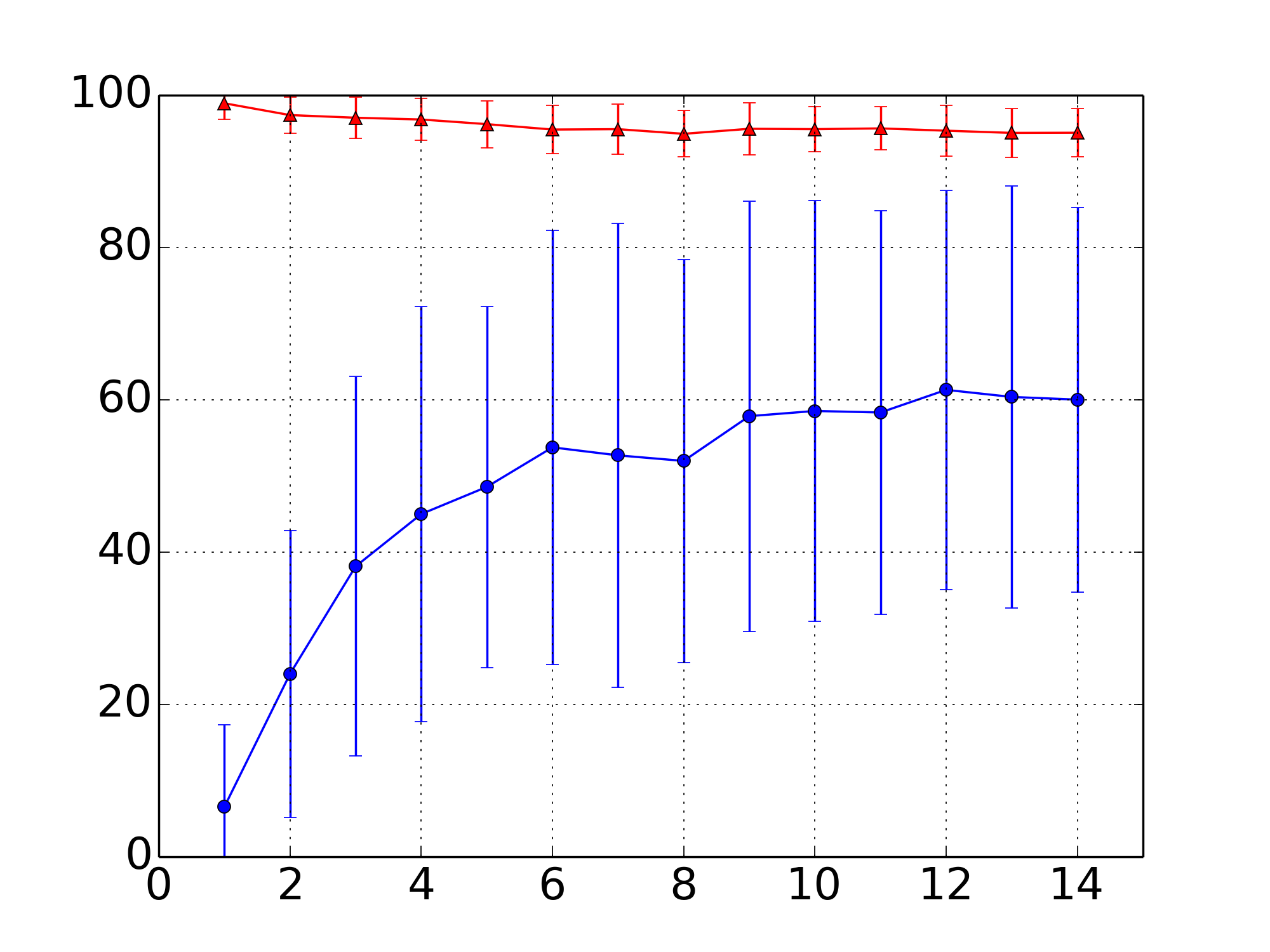}
  \includegraphics[width=.23\columnwidth, trim={1.22cm 0.82cm 2.0cm 1.1cm},clip]{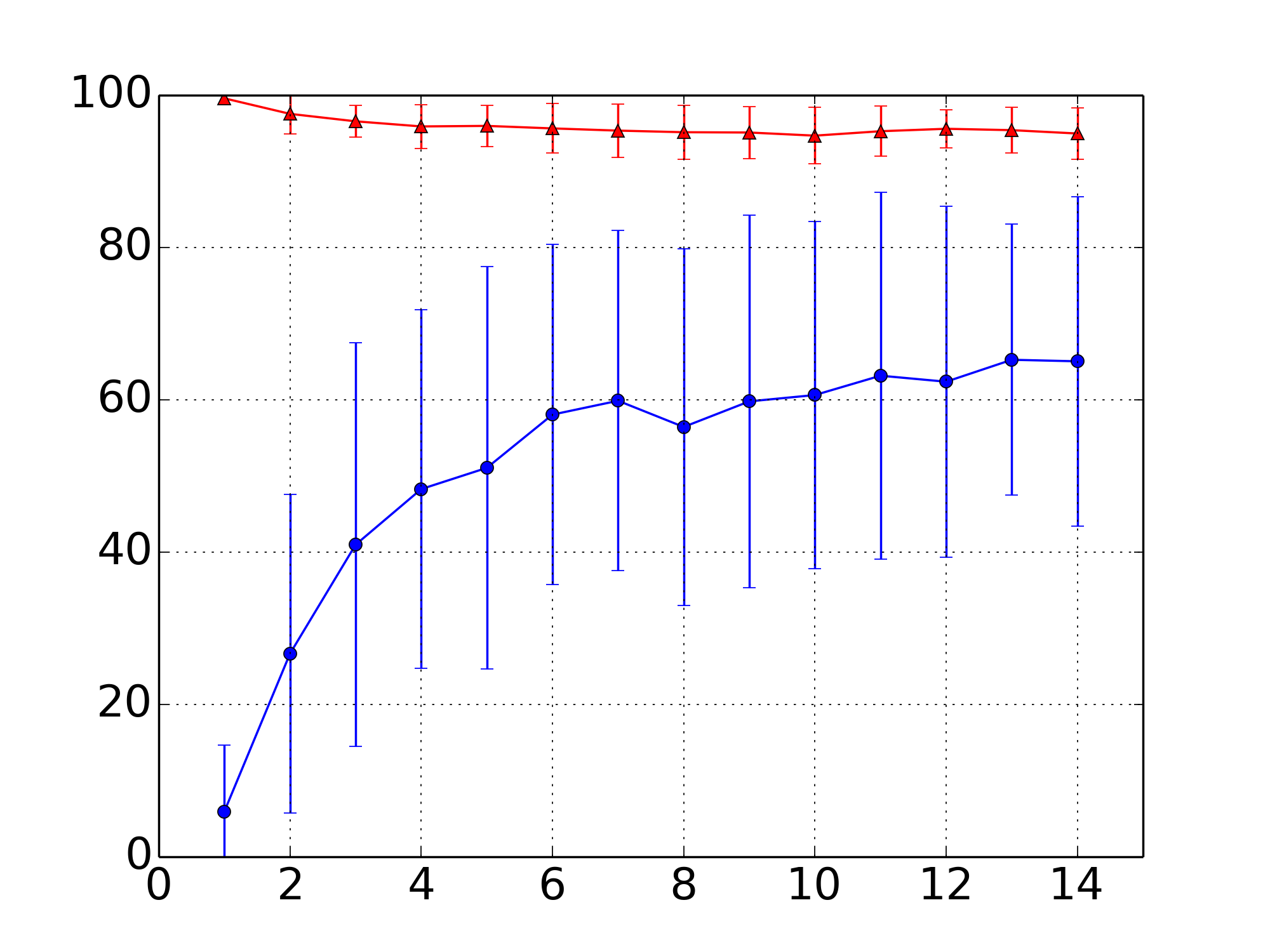}
  \includegraphics[width=.23\columnwidth, trim={1.22cm 0.82cm 2.0cm 1.1cm},clip]{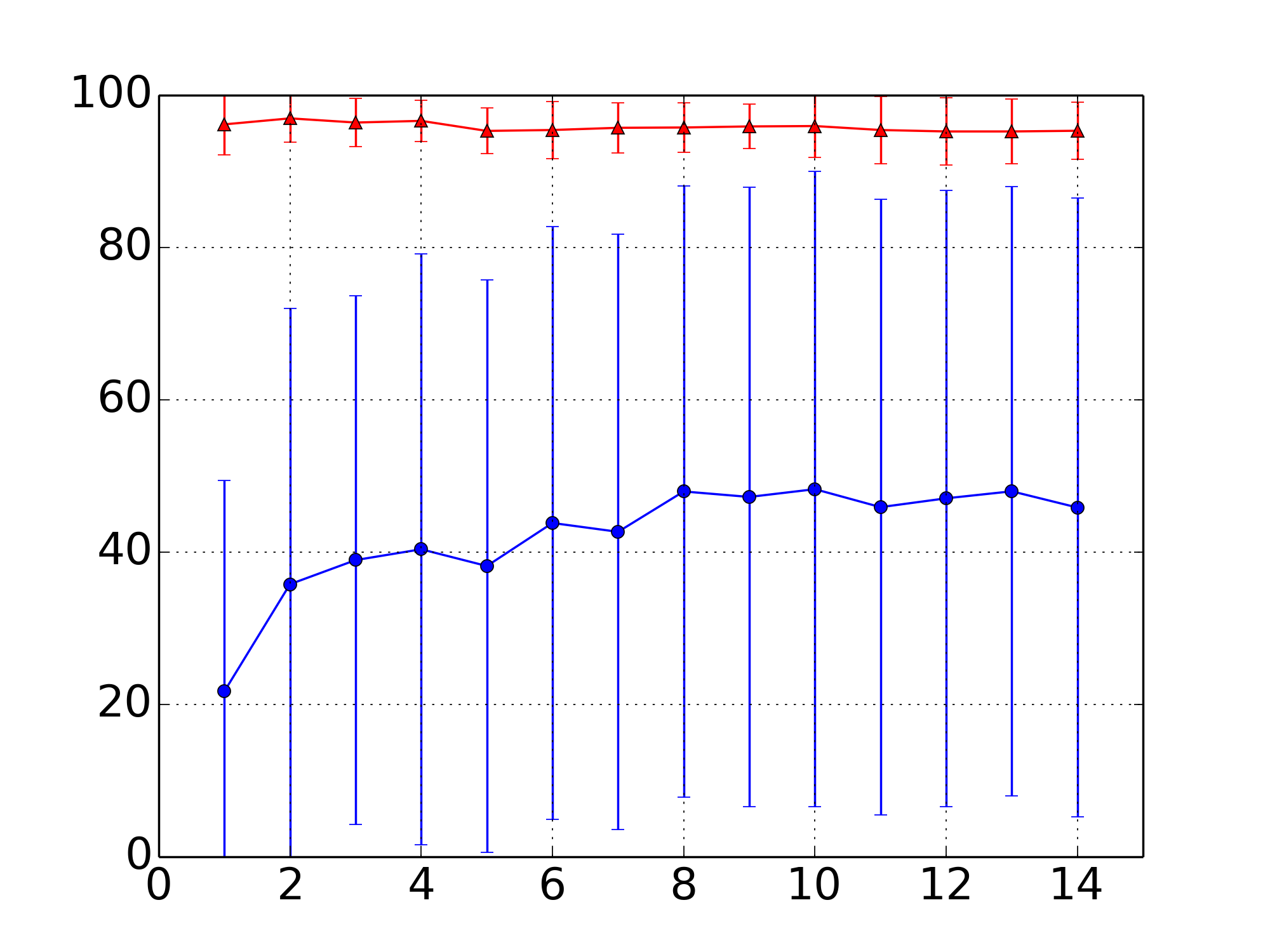}

  \includegraphics[width=.23\columnwidth, trim={1.22cm 0.82cm 2.0cm 1.1cm},clip]{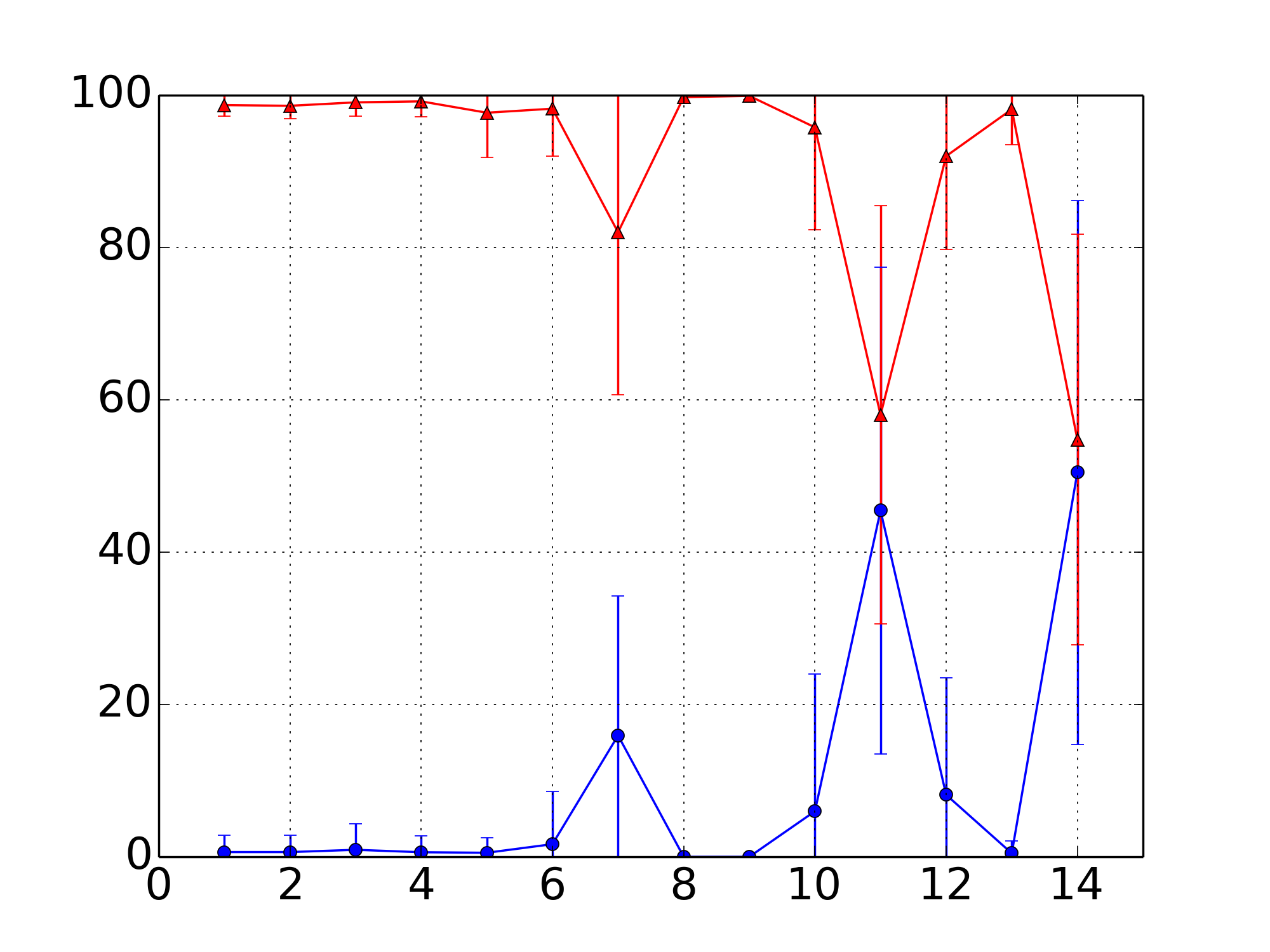}
  \includegraphics[width=.23\columnwidth, trim={1.22cm 0.82cm 2.0cm 1.1cm},clip]{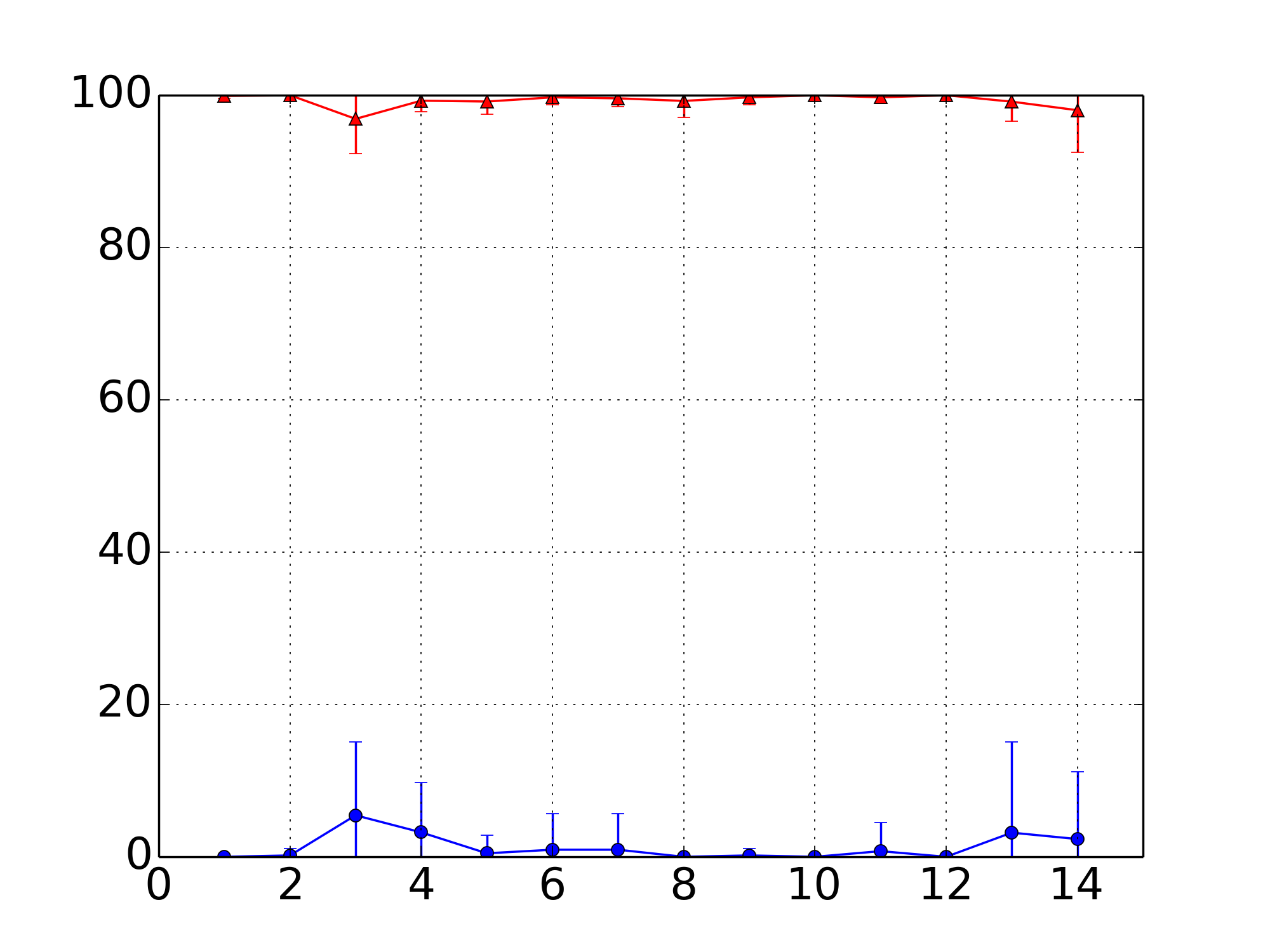}
  \includegraphics[width=.23\columnwidth, trim={1.22cm 0.82cm 2.0cm 1.1cm},clip]{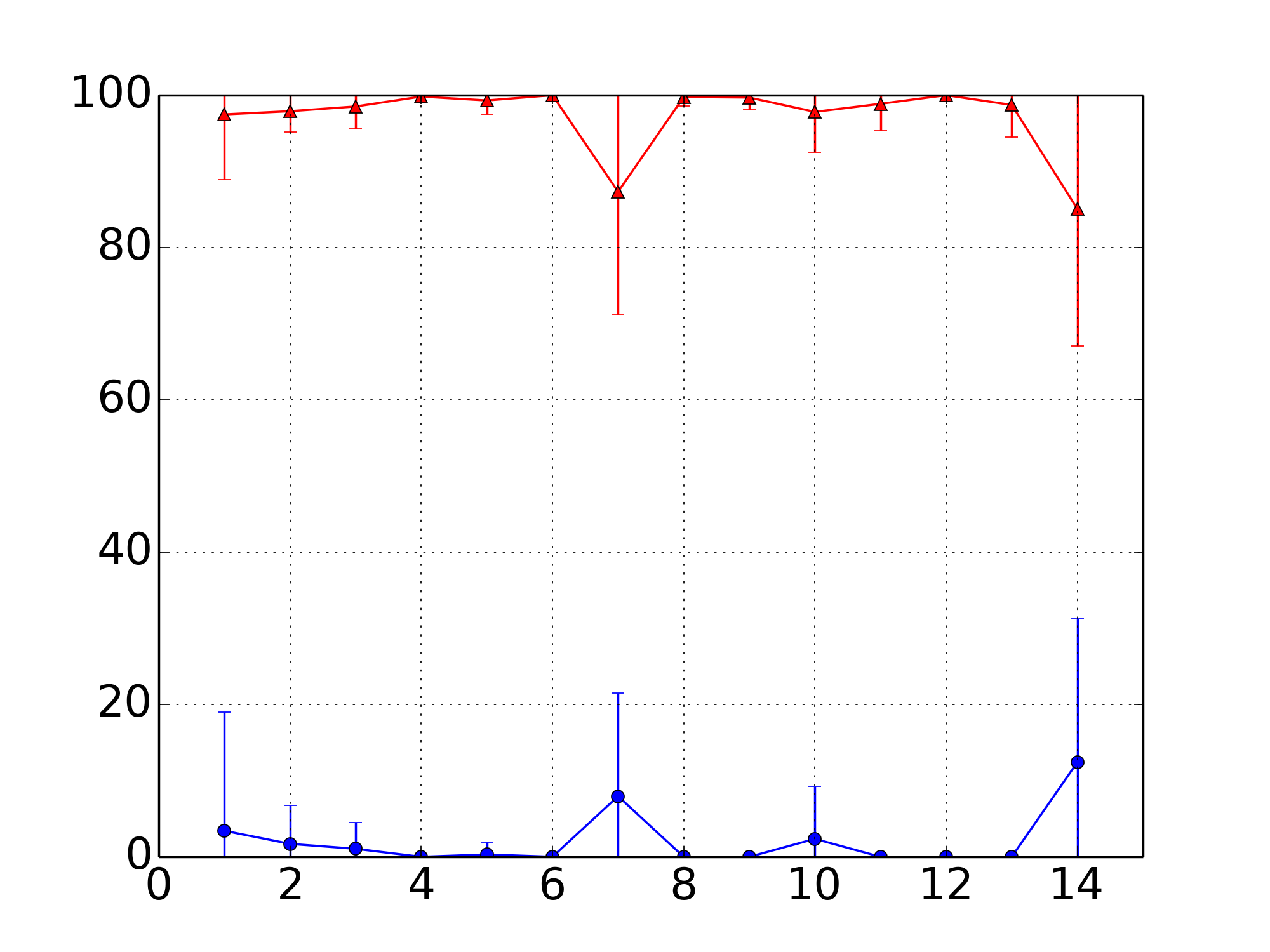}
  \includegraphics[width=.23\columnwidth, trim={1.22cm 0.82cm 2.0cm 1.1cm},clip]{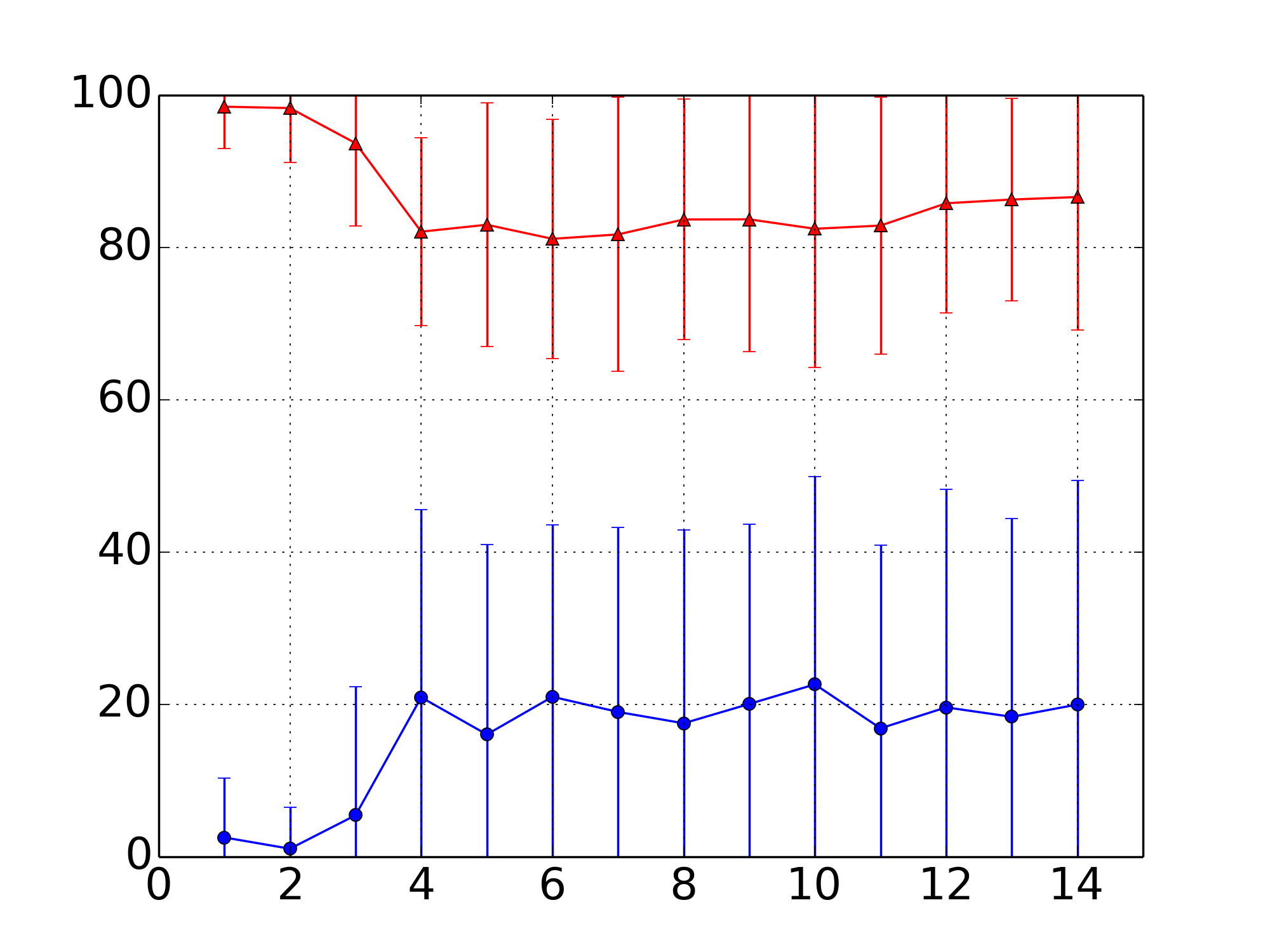}
\caption{{\myfont P2}: Sensitivity/specificity (blue/red) scores over the  \{positive, negative, wellbeing, stress\} targets by training on different time windows on the \textit{LOIOCV} (top) and \textit{LOUOCV} (bottom) setups, similar to \cite{canzian2015trajectories}.}
\label{fig:canzian_intra}
\end{figure}

The charts in Fig.~\ref{fig:canzian_intra} (top) show the results by following the \textit{LOIOCV} approach from Canzian and Musolesi \cite{canzian2015trajectories}. The pattern that these metrics take is consistent and quite similar to the original paper: specificity remains at high values, while sensitivity increases as we increase the time window from which we extract our features. The charts on the bottom in Fig. \ref{fig:canzian_intra} show the corresponding results in the \textit{LOUOCV} setting. Here, such a generalisation is not feasible, since the increases in sensitivity are accompanied by sharp drops in the specificity scores. 

The arising issue though lies in the \textit{LOIOCV} setting. By training and testing on the same days (for $T_{HIST}>1$), the kernel matrix takes high values for cells which are highly correlated with respect to time, making the evaluation of the contribution of the features difficult. To support this statement, we train the same model under \textit{LOIOCV}, using only on the mood form completion date (Unix epoch) as a feature. The results are  very similar to those achieved by training on $T_{HIST}=14$ (see Table \ref{tab:canzian}). We also include the results of another na\"ive classifier (\texttt{LAST}), predicting always the last observed score in the training set, which again achieves similar results. The clearest demonstration of the problem though is by comparing the results of the \texttt{RAND} against the \texttt{FEAT} classifier, which shows that under the proposed evaluation setup we can achieve similar performance if we replace our inputs with random data, clearly demonstrating the temporal bias that can lead to over-optimistic results, even in the \textit{LOIOCV} setting.

\begin{table}
\begin{center}
\resizebox{.65\textwidth}{!}{
\begin{tabular}{ |l||c|c||c|c||c|c||c|c|}
 \hline
& \multicolumn{2}{c||}{positive} & \multicolumn{2}{c||}{negative} & \multicolumn{2}{c||}{wellbeing}&
\multicolumn{2}{c|}{stress}\\
 \hline
& sens & spec & sens & spec & sens & spec & sens & spec \\
\hline
\texttt{FEAT} & 64.02 & 95.23 & 60.03 & 95.07 & 65.06 & 94.97& 45.86 & 95.32 \\
\texttt{DATE} & 59.68 & 95.92 & 62.75 & 95.19 & 63.29 & 95.47& 46.99 & 95.17\\
\texttt{LAST} & 67.37 & 94.12 & 69.08 & 94.09 & 66.05 & 93.42& 58.20 & 93.83 \\
\texttt{RAND}&64.22&95.17&60.88&95.60&64.87&95.09& 45.79&95.41 \\
\hline
\end{tabular}
}
\caption{{\myfont P2}: Performance (sensitivity/specificity) of the SVM classifier trained over $14$ days of smartphone/social media features (\texttt{FEAT}) compared against 3 na\"ive baselines.}
\label{tab:canzian}
\end{center}
\end{table}

\subsection{{\myfont P3}: Predicting Users}
\begin{figure}
\begin{floatrow}[2]
\ffigbox[1.172\FBwidth]{%
  \includegraphics[width=0.3\textwidth, trim={1.7cm 1.0cm 1.75cm 1.3cm},clip]{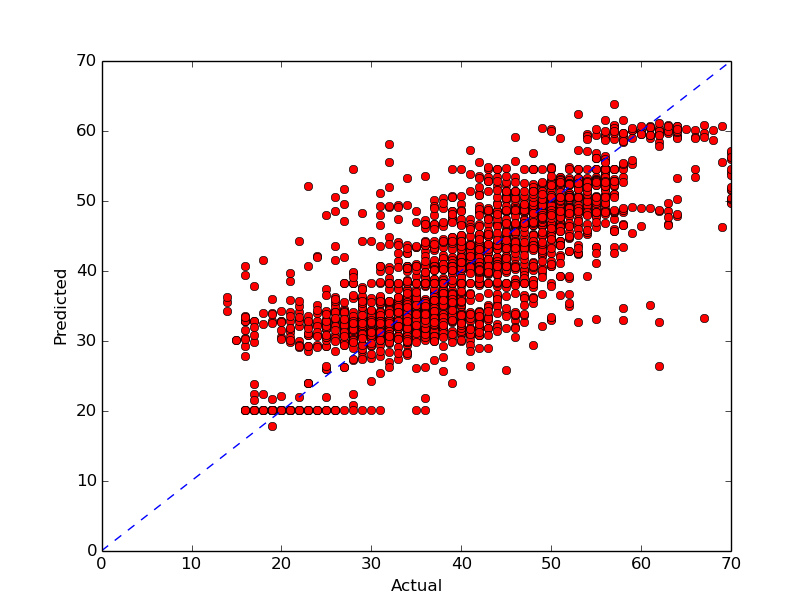}%
}{%
  \caption{{\myfont P3}: Actual vs predicted chart for the ``wellbeing'' target in \textit{LOIOCV}. The across-subjects $R^2$ is negative.}
  \label{fig:tsakalidisresults}%
}
\capbtabbox{%
\resizebox{0.62\textwidth}{!}{
\begin{tabular}{ |l||r|r||r|r||r|r||r|r|}
 \hline
& \multicolumn{2}{c||}{positive} & \multicolumn{2}{c||}{negative} & \multicolumn{2}{c||}{wellbeing}& 
\multicolumn{2}{c|}{stress} \\
 \hline
& $R^2$ & RMSE & $R^2$ &RMSE& $R^2$ & RMSE&$R^2$ & RMSE\\
\hline
MIXED$_+$ & 0.43 & 6.91 & 0.25 & 6.49 & 0.48 & 8.04 &0.02&1.03\\ 
MIXED$_-$ & 0.13 & 8.50  & 0.00  & 7.52 & 0.13 & 10.33 &0.03&1.03\\\hline
LOIOCV$_+$ & -0.03 & 5.20 & -0.04 & 5.05 & -0.03 & 6.03 &-0.08&0.91\\ 
LOIOCV$_-$ & -0.03 & 5.20 & -0.04 & 5.05 & -0.03 & 6.03 &-0.08&0.91\\\hline
LOUOCV$_+$ & -4.19 & 8.98 & -1.09 & 7.24 & -4.66 & 10.61 &-0.67&1.01\\
LOUOCV$_-$ & -4.38 & 8.98  & -1.41 & 7.23 & -4.62 & 10.62 &-0.69&1.02\\
\hline
\end{tabular}}}
{%
  \caption{{\myfont P3}: Results following the evaluation setup in \cite{tsakalidis2016combining} (\textit{MIXED}), along with the results obtained in the \textit{LOIOCV} and \textit{LOUOCV} settings with (+) and without (-) per-user input normalisation.}
  \label{tab:tsakalidisresults}%
}

\end{floatrow}
\end{figure}

\noindent \textbf{Experiment 1}:
Table~\ref{tab:tsakalidisresults} shows the results based on the evaluation setup of Tsakalidis et al. \cite{tsakalidis2016combining}. In the \textit{MIXED} cases, the pattern is consistent with \cite{tsakalidis2016combining}, indicating that normalising the features on a per-user basis yields better results, when dealing with sparse textual features (positive, negative, wellbeing targets). 
The explanation of this effect lies within the danger of predicting the user's identity instead of her mood scores. This is why the per-user normalisation does not have any effect for the stress target, since for that we are using dense features derived from smartphones: the vocabulary used by the subjects for the other targets is more indicative of their identity. In order to further support this statement, we trained the SVR model using only the one-hot encoded user id as a feature, without any textual features. Our results yielded $R^2$=\{$0.64$, $0.50$, $0.66$\} and $RMSE$=\{$5.50$, $5.32$, $6.50$\} for the \{positive, negative, wellbeing\} targets, clearly demonstrating the user bias in the \textit{MIXED} setting.

The RMSEs in \textit{LOIOCV} are the lowest, since different individuals exhibit different ranges of mental health scores. Nevertheless, $R^2$ is slightly negative, implying again that the average predictor for a single user provides a better estimate for her mental health score. Note that while the predictions across all individuals seem to be very accurate (see Fig.~\ref{fig:tsakalidisresults}), by separating them on a per-user basis, we end up with a negative $R^2$. 

In the unbiased \textit{LOUOCV} setting the results are, again, very poor. The reason for the high differences observed between the three settings is provided by the $R^2$ formula itself ($1 - (\sum_{i} (pred_i-y_i)^2)/(\sum_{i} (y_i-\bar{y})^2))$
. In the \textit{MIXED} case, we train and test on the same users, while $\bar{y}$ is calculated as the mean of the mood scores across all users, whereas in the \textit{LOIOCV}/\textit{LOUOCV} cases, $\bar{y}$ is calculated for every user separately. In \textit{MIXED}, by identifying who the user is, we have a rough estimate of her mood score, which is by itself a good predictor, if it is compared with the average predictor across all mood scores of all users. Thus, the effect of the features in this setting cannot be assessed with certainty.

\noindent \textbf{Experiment 2:}
Table~\ref{tab:jaques} displays our results based on Jaques et al. \cite{jaques2015predicting} (see section \ref{sec:tsakalidis}). The average accuracy on the \textit{``UNIQ''} setup is higher by $14\%$ compared to the majority classifier in \textit{MIXED}. The \textit{LOIOCV} setting also yields very promising results (mean accuracy: $81.17\%$). As in all previous cases, in \textit{LOUOCV} our models fail to outperform the majority classifier. A closer look at the \textit{LOIOCV} and \textit{MIXED} results though reveals the user bias issue that is responsible for the high accuracy. For example, $33\%$ of the users had all of their ``positive'' scores binned into one class, as these subjects were exhibiting higher (or lower) mental health scores throughout the experiment, whereas another $33\%$ of the subjects had $85\%$ of their instances classified into one class. By recognising the user, we can achieve high accuracy in the \textit{MIXED} setting; in the \textit{LOIOCV}, the majority classifier can also achieve at least $85\%$ accuracy for $18$/$27$ users. 

\begin{table}
\begin{center}
\resizebox{.8\textwidth}{!}{
\begin{tabular}{ |l||r|r||r|r||r|r||r|r|}
 \hline
& \multicolumn{2}{c||}{positive} & \multicolumn{2}{c||}{negative} & \multicolumn{2}{c||}{wellbeing}& 
\multicolumn{2}{c|}{stress}\\
 \hline
&UNIQ&PERS&UNIQ&PERS&UNIQ&PERS&UNIQ&PERS\\
\hline
MIXED &65.69&51.54&60.68&55.79&68.14&51.00&61.75&56.44\\
LOIOCV &78.22&51.79&84.86&53.63&88.06&52.89&73.54&55.35\\
LOUOCV &47.36&50.74&42.41&52.45&45.57&50.10&49.77&55.11\\
\hline
\end{tabular}
}
\caption{{\myfont P3}: Accuracy by following the evaluation setup in \cite{jaques2015predicting} (\textit{MIXED}), along with the results obtained in \textit{LOIOCV} \& \textit{LOUOCV}.}
\label{tab:jaques}
\end{center}
\end{table}

In the \textit{``PERS''} setup, we removed the user bias, by separating the two classes on a per-user basis. The results now drop heavily even in the two previously well-performing settings and can barely outperform the majority classifier. Note that the task in Experiment 2 is relatively easier, since we are trying to classify instances into two classes which are well-distinguished from each other from a psychological point of view. However, by removing the user bias, the contribution of the user-generated features to this task becomes once again unclear.

\section{Proposal for Future Directions}
Our results emphasize the difficulty of automatically predicting individuals' mental health scores in a real-world setting and demonstrate the dangers due to flaws in the experimental setup. Our findings do not imply that the presented issues will manifest themselves to the same degree in different datasets -- e.g., the danger of predicting the user in the \textit{MIXED} setting is higher when using the texts of 27 users rather than sensor-based features of more users \cite{bogomolov2014pervasive,bogomolov2013happiness,jaques2015predicting,servia2017mobile}. Nevertheless, it is crucial to establish appropriate evaluation settings to avoid providing false alarms to users, if our aim is to build systems that can be deployed in practice. To this end, we propose model building and evaluation under the following:

\begin{itemize}
    \item \textbf{LOUOCV}: By definition, training should be performed strictly on features and target data derived from a sample of users and tested on a completely new user, since using target data from the unseen user as features violates the independence hypothesis. A model trained in this setting should achieve consistently better results on the unseen user compared to the na\"ive (from a modelling perspective) model that always predicts his/her average score. 
    \item \textbf{LOIOCV}: By definition, the models trained under this setting should not violate the iid hypothesis. We have demonstrated that the temporal dependence between instances in the train and test set can provide over-optimistic results. A model trained on this setting should consistently outperform na\"ive, yet competitive, baseline methods, such as the last-entered mood score predictor, the user's average mood predictor and the auto-regressive model.  
\end{itemize}

Models that can be effectively applied in any of the above settings could revolutionise the mental health assessment process while  providing us in an unbiased setting with great insights on the types of behaviour that affect our mental well-being. On the other hand, positive results in the \textit{MIXED} setting cannot guarantee model performance in a real-world setting in either \textit{LOUOCV} or \textit{LOIOCV}, even if they are compared against the user average baseline \cite{demasi2017meaningless}.

\textbf{Transfer learning} approaches can provide significant help in the \textit{LOUOCV} setting. However, these assume that single-domain models have been effectively learned beforehand -- but all of our single-user (\textit{LOIOCV}) experiments provided negative results. Better feature engineering through \textbf{latent feature representations} may prove to be beneficial. While different users exhibit different behaviours, these behaviours may follow similar patterns in a latent space. Such representations have seen great success in recent years in the field of natural language processing \cite{mikolov2013distributed}, where the aim is to capture latent similarities between seemingly diverse concepts and represent every feature based on its context. Finally, working with \textbf{larger datasets} can help in providing more data to train on, but also in assessing the model's ability to generalise in a more realistic setting.

\section{Conclusion}

Assessing mental health with digital media is a task which could have great impact on 
monitoring of mental well-being and personalised health. In the current paper, we have followed past experimental settings to evaluate the contribution of various features to the task of automatically predicting different mental health indices of an individual. We find that under an unbiased, real-world setting, the performance of state-of-the-art models drops significantly, making the contribution of the features impossible to assess. Crucially, this holds for both cases of creating a model that can be applied in previously unobserved users (\textit{LOUOCV}) and a personalised model that is learned for every user individually (\textit{LOIOCV}). 

Our major goal for the future is to achieve positive results in the \textit{LOUOCV} setting. To overcome the problem of having only few instances from a diversely behaving small group of subjects, transfer learning techniques on latent feature representations could be beneficial. A successful model in this setting would not only provide us with insights on what types of behaviour affect mental state, but could also be employed in a real-world system without the danger of providing false alarms to its users.

\section*{Acknowledgements}
The current work was supported by the EPSRC through the University of Warwick's CDT in Urban Science and Progress (grant EP/L016400/1) and through The Alan Turing Institute (grant EP/N510129/1). We would like to thank the anonymous reviewers for their detailed feedback and the authors of the works that were analysed in our paper (N. Jaques, R. LiKamWa, M. Musolesi) for the fruitful discussions over several aspects of the presented challenges.

\bibliographystyle{splncs04}
\bibliography{refs}
\end{document}